\documentclass[prl,preprint,showpacs]{revtex4}

\usepackage{amsfonts}
\usepackage{amssymb}
\usepackage{graphicx}
\usepackage{pstricks}

\newcommand{\be}{\begin{equation}}
\newcommand{\bea}{\begin{eqnarray}}
\newcommand{\ee}{\end{equation}}
\newcommand{\eea}{\end{eqnarray}}
\newcommand{\bpi}{\begin{picture}}
\newcommand{\bce}{\begin{center}}
\newcommand{\epi}{\end{picture}}
\newcommand{\ece}{\end{center}}

\newcommand{\ksm}{k\hspace{-0.235cm}/}

\newcommand{\psm}{p\hspace{-0.13cm}/}
\newcommand{\psmp}{p\hspace{-0.16cm}/}

\def\chic#1{{\scriptscriptstyle #1}}
\def\r#1{(\ref{#1})}

\begin{document}

\title{The Pinch Technique to All Orders}
\date{August 21, 2002}

\author{Daniele Binosi}
\author{Joannis Papavassiliou}
\affiliation{Departamento de F\'\i sica Te\'orica and IFIC, Centro Mixto, 
Universidad de Valencia-CSIC,
E-46100, Burjassot, Valencia, Spain}

\email{Daniele.Binosi@uv.es; Joannis.Papavassiliou@uv.es}

\begin{abstract}

The  generalization   of  the  pinch   technique  to  all   orders  in
perturbation  theory is  presented.  The  effective  Green's functions
constructed  with this  procedure  are  singled out  in  a unique  way
through     the     full     exploitation    of     the     underlying
Becchi-Rouet-Stora-Tyutin symmetry.  A simple all-order correspondence
between the  pinch technique  and the background  field method  in the
Feynman gauge is established.

\end{abstract}

\pacs{11.15.Bt,12.38.Bx,14.70.Dj}

\preprint{FTUV-02-0821}
\preprint{IFIC-02-19}

\maketitle

It  is  well-known  that,   to  any  finite  order,  the  conventional
perturbative  expansion   gives  rise  to   expressions  for  physical
amplitudes  which  are endowed  with  crucial properties.   $S$-matrix
elements, for example, are  independent of the gauge-fixing scheme and
parameters  chosen to  quantize the  theory, they  are gauge-invariant
(current   conservation),   they    are   unitary   (conservation   of
probability), and  well behaved at  high energies. However,  the above
properties are  in general not  reflected by the  individual off-shell
Green's functions, which are the building blocks of the aforementioned
perturbative  expansion.   The   latter  depend  on  the  gauge-fixing
parameters  in a  complicated way  ,  grow much  faster than  physical
amplitudes  at  high  energies,  and  display  unphysical  thresholds.
Evidently,  when  combining unphysical  Green's  functions  to form  a
physical amplitude,  subtle field-theoretical mechanisms  are at work,
which enforce non-trivial cancellations among them at any given order.

There are  considerable conceptual and  phenomenological advantages in
reformulating the perturbative expansion in terms of off-shell Green's
functions which display manifestly the same properties as the physical
amplitudes.  To  begin with, the sharp  difference between observables
and  Green's functions  suggests a  great  deal of  redundancy in  the
conventional diagrammatic formulation of  gauge theories, in the sense
that extensive underlying cancellations beg to be made manifest and be
explicitly  exploited  as  early  within a  calculation  as  possible.
Implementing these  cancellations at an  early stage not  only renders
the book-keeping aspects more tractable \cite{Feng:1995vg}, but allows
for  theoretically   safe  reorganizations  or   resummations  of  the
perturbative series.  For  example, identifying and Dyson-resuming the
correct   sub-set  of  propagator-like   corrections  gives   rise  to
physically         meaningful         Born-improved         amplitudes
\cite{Papavassiliou:1995fq}.  In  addition, the generalization  into a
non-Abelian  context  of  the  characteristic properties  of  the  QED
effective charge,  has a  wide range of  phenomenological applications
\cite{Brodsky:2001wx}.   Finally,  $n$-point
functions  free  of  unphysical  artifacts  could serve,  at  least  in
principle, as  the new  building blocks of  manifestly gauge-invariant
Schwinger-Dyson equations \cite{Cornwall:1982zr}.

It would  clearly be preferable to enforce  the relevant cancellations
already  at the  level of  the functional  path-integral  defining the
theory,  and obtain directly  from it  the desired  Green's functions;
this is however  beyond our powers at the moment.   On the other hand,
there exists  a {\it diagrammatic} method, called  the pinch technique
(PT)   \cite{Cornwall:1982zr,Cornwall:1989gv},   which  systematically
exploits  the  symmetries built  into  physical  observables, such  as
$S$-matrix  elements, in order  to construct  off-shell sub-amplitudes
that are  kinematically akin  to conventional Green's  functions, but,
unlike the  latter, are also  endowed with desirable  properties.  The
basic  observation, which essentially  defines the  PT, is  that there
exists  a  fundamental  cancellation  between sets  of  diagrams  with
different kinematic  properties, such as  self-energies, vertices, and
boxes.     This   cancellation   is    driven   by    the   underlying
Becchi-Rouet-Stora-Tyutin   symmetry   \cite{Becchi:1976nq},  and   is
triggered when longitudinal momenta  circulating inside vertex and box
diagrams  generate  (by   ``pinching''  out  internal  fermion  lines)
propagator-like  terms.   The latter  are  reassigned to  conventional
self-energy  graphs  in  order  to  give rise  to  the  aforementioned
gauge-invariant  effective Green's  functions.  In  its  original one-
\cite{Cornwall:1982zr,Cornwall:1989gv}           and          two-loop
\cite{Papavassiliou:2000az}  application,  the PT  boils  down to  the
study of  the kinematic rearrangements produced  into {\it individual}
Feynman  diagrams  when  elementary  tree-level  Ward  identities  
(WIs) are triggered.

One of the most pressing questions  in this context is whether one can
extend the  PT algorithm  to all orders  in perturbation  theory, thus
achieving the systematic construction of effective $n$-point functions
displaying the aforementioned  characteristic features.  To accomplish
this it is clear that one needs to abandon algebraic operations inside
individual Feynman graphs, and resort  to a more formal procedure.  In
this Letter  we will  show that the  PT algorithm can  be successfully
generalized to  {\it all orders}  in perturbation theory,  through the
collective  treatment  of entire  sets  of  diagrams.
This is  accomplished through the judicious use  of the Slavnov-Taylor
identity  (STI) \cite{Slavnov:1972fg} satisfied  by a  special Green's
function,  which  serves  as  a  common kernel  to  all  higher  order
self-energy and vertex diagrams.

We  will consider  for  concreteness the  $S$-matrix
element  for the  quark--anti-quark elastic  scattering  process $
q(r_1)\bar q(r_2)\to q(p_1)\bar q(p_2)$ in  QCD. We set $q=r_2-r_1=p_2-p_1$,
with $s=q^2$ the  square of the momentum transfer. The  longitudinal
momenta  responsible  for the aforementioned kinematical 
rearrangements stem either from the bare gluon propagators or from the
pinching  part $\Gamma^{{\rm  P}}_{\alpha\mu\nu}(q,k_1,k_2)$ appearing
in  the  characteristic  decomposition  of the  elementary  tree-level
three-gluon   vertex $\Gamma^{eab,[0]}_{\alpha\mu\nu}=
gf^{eab}\Gamma^{[0]}_{\alpha\mu\nu}$ into 
\cite{Cornwall:1982zr}    
\bea
\Gamma^{[0]}_{\alpha\mu\nu}(q,k_1,k_2)&=&  \Gamma^{{\rm
F}}_{\alpha\mu\nu}(q,k_1,k_2)+ \Gamma^{{\rm
P}}_{\alpha\mu\nu}(q,k_1,k_2),        \nonumber\\ 
\Gamma^{{\rm
F}}_{\alpha\mu\nu}(q,k_1,k_2)&=&(k_1-k_2)_\alpha  g_{\mu\nu}+2q_\nu
g_{\alpha\mu}-2q_\mu g_{\alpha\nu},\nonumber    \\    
\Gamma^{{\rm
P}}_{\alpha\mu\nu}(q,k_1,k_2)&=&k_{2\nu}g_{\alpha\mu}-
k_{1\mu}g_{\alpha\nu}.
\label{PTDEC}
\eea 
The above  decomposition is  to be  carried out  to ``external''
three-gluon  vertices  only, {\it i.e.},   the  vertices  where the  physical
momentum transfer $q$ is entering \cite{Papavassiliou:2000az}.   In
what follows we work  
in the renormalizable  Feynman gauge (RFG);  this choice
eliminates  the longitudinal  momenta from  the bare  propagators, and
allows  us  to focus  our  attention on  the  all-order  study of  the
longitudinal      momenta      originating     from      $\Gamma^{{\rm
P}}_{\alpha\mu\nu}$.
\begin{figure}[!t]
\includegraphics[width=6.0cm]{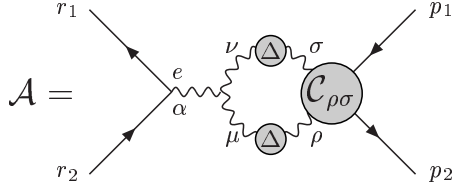}
\caption{\label{fig:0}   The    subset   of   the    graphs   of   the
quark--anti-quark  elastic scattering process  which will  receive the
action of the longitudinal momenta stemming from $\Gamma^{{\rm
P}}$. Here $\Delta$ represents the full gluon propagator.}
\end{figure}
We will denote by
$\cal{A}$ the  subset of the graphs  which will receive  the action of
the     longitudinal    momenta     stemming     from    $\Gamma^{{\rm
P}}_{\alpha\mu\nu}(q,k_1,k_2)$ (see Fig.\ref{fig:0}). We have that
\be 
{\cal A} = ig^2\bar  u(r_1) 
\frac{\lambda^e}{2} \gamma_{\alpha} v(r_2) 
f^{eab} \Gamma^{{\rm P}\!,\,\alpha\mu\nu}(q,k_1,k_2)
{\cal T}^{ab}_{\mu\nu}(k_1,k_2,p_1,p_2),
\ee
where   $\lambda^e$   are   the   Gell-Mann   matrices,   and   ${\cal
T}_{\mu\nu}^{ab}$     is    the     sub-amplitude    $g_{\mu}^{a}(k_1)
g_{\nu}^{b}(k_2)\to q(p_1)\bar q(p_2)$, with the gluons {\it off-shell}
and    the   fermions    on-shell;    for   the    latter   $\left.\bar
v(p_2)S^{-1}(p_2)\right|_{\psm_2=m}=
\left.S^{-1}(p_1)u(p_1)\right|_{\psm_1=m}=0$,  where  $S(p)$  is  the
(full) quark propagator.  In terms of Green's functions we have
\be
{\cal T}_{\mu\nu}^{ab} = 
\bar  v(p_2)\!\left[{\cal C}^{ab}_{\rho\sigma}(k_1,k_2,p_1,p_2)
\Delta^{\rho}_{\mu}(k_1)\Delta^{\sigma}_{\nu}(k_2)\right]\!u(p_1).
\label{sgf}
\ee
Clearly,  there  is  an  equal contribution from the  $\Gamma^{{\rm
P}}$ situated on the right hand-side of ${\cal T}$.

\begin{figure}[!t]
\includegraphics[width=11.5cm]{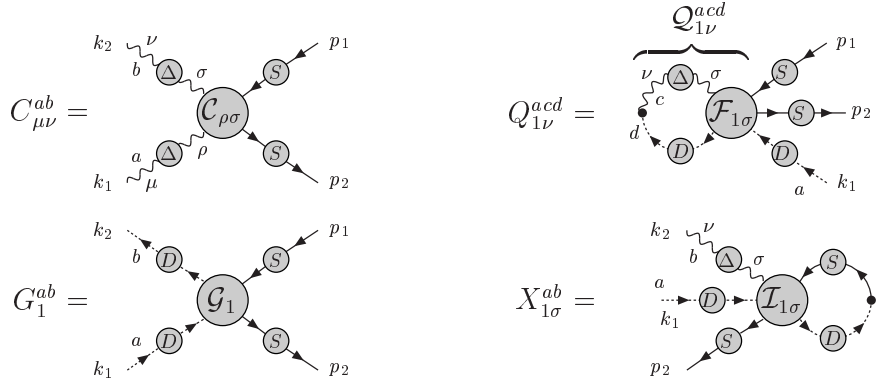}
\caption{\label{fig:2} Diagrammatic representation of the
Green's function appearing in the STI of Eq.\r{BasSTI}. Here $D$ and
$S$ represent the full ghost and fermion  propagators respectively.}
\end{figure}

Let us focus on the STI satisfied by the amplitude 
${\cal T}_{\mu\nu}^{ab}$; it reads
\bea
k_1^\mu C_{\mu\nu}^{ab}+k_{2\nu}G_1^{ab}-igf^{bcd}
Q_{1\nu}^{acd}
-gX_{1\nu}^{ab}
+g\bar X_{1\nu}^{ab}=0, 
\label{BasSTI}
\eea  
where the Green's function appearing in it are defined in 
Fig.\ref{fig:2}. The  terms
$X_{1\nu}$ and   $\bar  X_{1\nu}$ 
die  on-shell, since  they are  missing one
fermion  propagator.  
Thus,  we
arrive at the on-shell STI for ${\cal T}_{\mu\nu}^{ab}$
\be
k_1^\mu {\cal T}_{\mu\nu}^{ab}=
{\cal S}^{ab}_{1\nu}, 
\label{onshSTI}
\ee
with
\bea
{\cal S}^{ab}_{1\nu}\!\!\!&=&\!
\bar  v(p_2)\left[igf^{bcd}{\cal
Q}_{1\nu}^{acd}(k_1,k_2,p_1,p_2)D(k_1)\right.\nonumber \\
&-&\left.
k_{2\nu}{\cal
G}_1^{ab}(k_1,k_2,p_1,p_2)D(k_1)D(k_2)\right]u(p_1), 
\label{onshdef}
\eea
where we have defined 
$Q_{1\nu}^{acd}(k_1,k_2,p_1,p_2)={\cal Q}_{1\nu}^{acd}(k_1,k_2,p_1,p_2)
D(k_1)S(p_1)S(p_2)$.

\begin{figure}[!t]
\includegraphics[width=7.5cm]{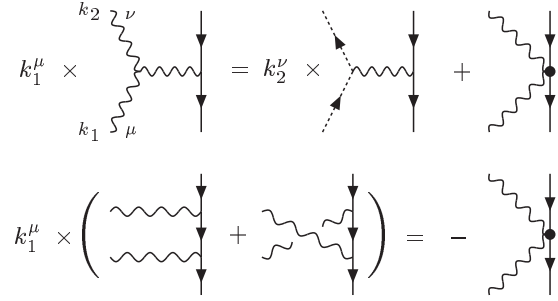}
\caption{\label{fig:4} Diagrammatic representation of the tree-level
inequality of Eq.\r{INEQ}.}
\end{figure}

In   perturbation   theory  both   ${\cal   T}^{ab}_{\mu\nu}$  and   ${\cal
S}^{ab}_{1\nu}$ are given by  Feynman diagrams, which can be separated
into  distinct classes,  depending on  their kinematic  dependence and
their geometrical properties.  Graphs which do not contain information
about  the  kinematical  details   of  the  incoming  test-quarks  are
self-energy graphs,  whereas those which  display a dependence  on the
test quarks are vertex graphs. The  former depend only on the variable
$s$,  whereas the latter  on both  $s$ and  the mass  $m$ of  the test
quarks; equivalently,  we  will  refer to  them  as $s$-channel  or
$t$-channel  graphs,   respectively.   In  addition   to  the  $s$-$t$
decomposition,  Feynman diagrams  can be  separated  into one-particle
irreducible (1PI) and one-particle  reducible (1PR) ones.  The crucial
point is  that the action of  the momenta $k_1^\mu$  or $k_2^\nu$ on
${\cal T}^{ab}_{\mu\nu}$  does {\it not} respect, in  general, the original
$s$-$t$ and 1PI-1PR separation furnished by the Feynman diagrams (see
third paper of~\cite{Papavassiliou:1995fq}).
In other
words, even though Eq.(\ref{onshSTI}) holds for the 
entire amplitude, 
it is not true for the individual sub-amplitudes, 
{\it i.e.},
\be
k_1^\mu \left[{\cal T}^{ab}_{\mu\nu}\right]_{x,{\rm \chic{Y}}} \neq  
\left[{\cal S}^{ab}_{1\nu}\right]_{x, {\rm \chic{Y}}},\qquad x=s,t; \quad
{\rm Y=I,R},
\label{INEQ}
\ee
where I (respectively R) indicates the one-particle {\it irreducible}
(respectively {\it reducible}) parts of the amplitude involved.
Evidently,   whereas   the   characterization   of   graphs   as
propagator- and  vertex-like is  unambiguous  in  the absence  of
longitudinal momenta  ({\it e.g.}, in a scalar theory),  their presence tends
to mix propagator- and  vertex-like graphs.  Similarly, 1PR graphs are
effectively converted into 1PI ones (the opposite cannot happen).  The
reason  for  the  inequality   of  Eq.(\ref{INEQ})  are  precisely  the
propagator-like  terms, such  as  those encountered  in  the one-  and
two-loop calculations; they have the characteristic feature that, when
depicted  by means  of Feynman  diagrams contain  unphysical vertices,
{\it   i.e.},  vertices   which   do  not   exist   in  the   original
Lagrangian (Fig.\ref{fig:4}).  All such diagrams  cancel {\it
diagrammatically} against  
each other.  Thus,  
after the aforementioned rearrangements have taken place, for the 
$t$-channel irreducible part of the amplitude we will have the equality
\be 
\left[k_1^\mu  {\cal
T}^{ab}_{\mu\nu}\right]_{t,{\rm \chic{I}}}^{\chic{\rm  PT}}  \equiv
\left[{\cal   S}^{ab}_{1\nu}\right]_{t,{\rm \chic{I}}}.
\label{EQPT}
\ee  

Eq.(\ref{EQPT}) merits  particular  attention, because  it is  of
central importance  for the  generalization of the  PT to  all orders.
The  superscript  ``PT''  on  the  left  hand-side  denotes  that  the
corresponding amplitude must  be rearranged following the well-defined
PT   algorithm,  as   it  has   been  explained   in   the  literature
\cite{Papavassiliou:2000az}.  In  particular,   one  tracks  down  the
rearrangments  induced  when  the  action  of  (virtual)  longitudinal
momenta ($k$) on the bare vertices of diagrams trigger elementary WIs.
Eventually  a WI  of the  form $  k_{\mu}\gamma^{\mu} =  S^{-1}(\ksm +
\psmp)-  \,S^{-1}(\psmp)$ will  give rise  to propagator-like  parts, by
removing (pinching out) the internal bare fermion propagator $S(\ksm +
\psmp)$.  Depending on the  topology of the diagram under consideration
this last WI may be activated  immediately, or as the final outcome of
a  sequential  triggering  of  intermediate  WIs.  We  emphasize  that,
in order to preserve the special 
unitarity and analyticity properties of the PT Green's functions, 
``internal''  three-gluon vertices  should not  pinch, nor  should one
carry out sub-integrations \cite{Papavassiliou:2000az}.  

The non-trivial step for generalizing the
PT to all  orders is then the following: Instead  of going through the
arduous task  of manipulating  the left hand-side  of Eq.(\ref{EQPT}),
following the aforementioned rules, in order to determine the pinching
parts  and explicitly  enforce  their cancellation,  use directly  the
right-hand  side,  which already  contains  the  answer!  Indeed,  the
right-hand side involves  only conventional (ghost) Green's functions,
expressed  in terms  of normal  Feynman  rules, with  no reference  to
unphysical vertices.  That this  must be so  follows from the  same PT
rules  mentioned above:  due to  the absence  of  external three-gluon
vertices  the right-hand  side cannot  be pinched  further,  i.e.  its
separation  into propagator-  and vertex-like  graphs  is unambiguous,
since  there is  no possibility  (without violating  the PT  rules) to
obtain further  mixing.  Thus, the right-hand  side of Eq.(\ref{EQPT})
serves as a practical definition of the PT to all orders.

After these  observations, we  proceed to the  PT construction  to all
orders. Once  the effective Green's functions have  been derived, they
will be  compared to the  corresponding Green's functions  obtained in
the Feynman  gauge of  the background field method (BFG for short) 
in
order  to establish  whether the  known 
correspondence  persists  to all
orders; as we will see, this is indeed the case
(for an extended list of related references see \cite{Binosi:2002ez}).

To  begin with,  it is  immediate  to recognize  that in  the RFG  box
diagrams of arbitrary order $n$,  to be denoted by $B^{[n]}$, coincide
with the PT boxes ${\widehat B}^{[n]}$, since all three-gluon vertices
are  ``internal'',  {\it  i.e.},   they  do  not  provide  longitudinal
momenta. Thus, they coincide  with the BFG boxes, $\tilde{B}^{[n]}$,
{\it  i.e.}, $  {\widehat B}^{[n]}  = B^{[n]}  =  \tilde{B}^{[n]}$ for
every~$n$.

\begin{figure}[!t]
\includegraphics[width=10.5cm]{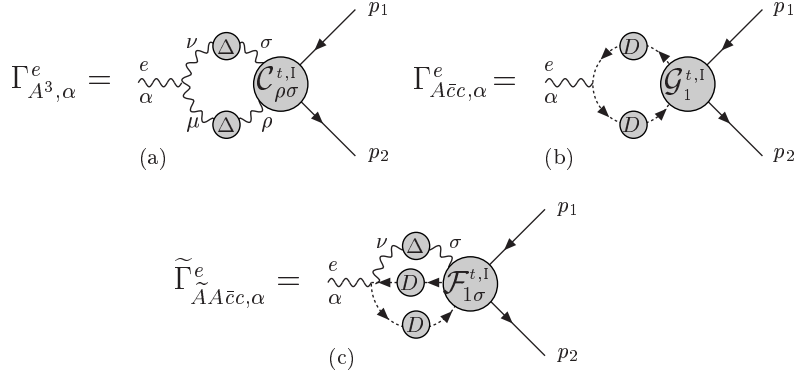}
\caption{\label{fig:1} The Green's functions identified in the
construction of the all order PT vertex
$\widehat\Gamma^{e}_\alpha$. The Green's functions 
(b) and (c) receive a contribution from
similar terms with the ghost arrows reversed (not shown).}
\end{figure}

We  then  continue with  the  construction  of  the  1PI
PT    gluon-quark--anti-quark     vertex    $\widehat
\Gamma^e_\alpha$.  We start from  the corresponding vertex in the RFG,
to be  denoted by  $\Gamma^e_\alpha$, and focus  only on the  class of
vertex diagrams containing an  {\it external} bare three-gluon vertex;
we   will   denote    this   subset   by   $\Gamma^{e}_{\!A^3,\alpha}$
[Fig.\ref{fig:1}(a)].   All  other  types  of graphs  contributing  to
$\Gamma^e_\alpha$ are inert  as far as the PT  procedure is concerned,
because     they      do     not     furnish      pinching     momenta
\cite{Papavassiliou:2000az}.  The next step is to carry out the vertex
decomposition of  Eq.(\ref{PTDEC}) to the  external three-gluon vertex
$\Gamma^{eab,[0]}_{\alpha\mu\nu}$             appearing             in
$\Gamma^{e}_{\!A^3,\alpha}$.    This  will   result  in   the  obvious
separation       $\Gamma^{e}_{\!A^3,\alpha}       =       \Gamma^{{\rm
F}\!,\,e}_{\!A^3,\alpha} +  \Gamma^{{\rm P}\!,\,e}_{\!A^3,\alpha}$. The
part $\Gamma^{{\rm  F}\!,\,e}_{\!A^3,\alpha}$ is also  inert, and will
be left untouched.  Thus, the  only quantity to be further manipulated
is $\Gamma^{{\rm P}\!,\,e}_{\!A^3,\alpha}$;  it reads 
\be 
\Gamma^{{\rm
P\!,\,}e}_{\!A^3,\alpha}=gf^{eba}\int \left[(k-q)^\mu
g^\nu_\alpha+k^\nu g^\mu_\alpha\right] \left[{\cal
T}_{\mu\nu}^{ab}\right]_{t,{\rm I}},   
\ee   
where    $\int   \equiv
\mu^{2\varepsilon}  \int  d^dk/(2\pi)^d$,  $d=D-2\varepsilon$,  
$D$ is the space-time dimension,
and
$\mu$ is the 't Hooft mass.  Following the discussion presented above,
the  pinching   action  amounts  to  the   replacement  $k^\nu  [{\cal
T}_{\mu\nu}^{ab}]_{t,{\rm       I}}       \to       [k^\nu       {\cal
T}_{\mu\nu}^{ab}]_{t,{\rm          I}}          =          \left[{\cal
S}_{2\mu}^{ab}(-k+q,k)\right]_{t,{\rm  I}}$ and  similarly  for the
term coming from the momentum 
$(k-q)^\mu$, {\it i.e.},  $[(k-q)^\mu       {\cal
T}_{\mu\nu}^{ab}]_{t,{\rm          I}}          =          -\left[{\cal
S}_{1\nu}^{ab}(-k+q,k)\right]_{t,{\rm  I}}$,  or,    equivalently,   
\be    
\Gamma^{{\rm 
P\!,\,}e}_{\!A^3,\alpha}(q) \rightarrow  gf^{eba} \int \bigg( [{\cal
S}_{2\alpha}^{ab}]_{t,{\rm  I}}  -  [{\cal  S}_{1\alpha}^{ab}]_{t,{\rm
I}}\bigg) .
\label{PTvertex}
\ee
At   this  point  the   construction  of   the  effective   PT  vertex
$\widehat\Gamma^{e}_{\alpha}$ has been  completed.  The next important
point is to study the connection between $\widehat\Gamma^{e}_{\alpha}$
and            the         vertex
$\widetilde\Gamma^{e}_{\alpha}$  in  the BFG.   To  begin with,  all
``inert'' terms contained  in the original $\Gamma^{e}_{\alpha}$ carry
over  to the same  sub-groups of  graphs obtained  in the  BFG; most
notably, the $\Gamma^{{\rm  F}\!,\,e}_{\!A^3,\alpha}$ is precisely the
$\widetilde\Gamma^{e}_{\!\widetilde{A}A^2,\alpha}$            part
of 
$\widetilde\Gamma^{e}_{\alpha}$,   where    $\widetilde{A}$   is   the
background   gluon.   The  only   exception  are   the  ghost-diagrams
contributing to $\Gamma^{e}_{\alpha}$ [Fig.\ref{fig:1}(b)]; the latter
do {\it  not} coincide with  the corresponding ghost  contributions in
the BFG.

The  important step is  to recognize  that the  BFG ghost  sector is
provided  precisely  by combining  the  RFG  ghosts  with the
right-hand side  of 
Eq.(\ref{EQPT}).  Specifically,  one arrives  at both  the {\it  symmetric}
vertex $\widetilde\Gamma_{\widetilde{A}\bar  c c}^{e}$, characteristic
of       the      BFG,      as       well      as       at    the
four-particle ghost vertex
$\widetilde\Gamma_{\!\widetilde{A} A\bar  c c}^{e}$, which  is totally
absent in the conventional formalism [Fig.\ref{fig:1}(c)].  Indeed we
find (omitting the spinors)
\bea
\int\!
\left[{\cal S}_{1\alpha}^{ab}\right]_{t,{\rm
I}}&=&\int\!D(-k+q)\left\{-k_\alpha\left[
{\cal G}_1^{ab}(-k+q,k)\right]_{t,{\rm
I}}D(k)\right. \nonumber \\
&+ &\left. igf^{bcd}
\left[
{\cal Q}_{1\alpha}^{acd}(-k+q,k)\right]_{t,{\rm
I}}\right\}.
\eea
A similar equation, in which we have to trade the ${\cal G}_1^{ab}$
and ${\cal Q}_{1\alpha}^{acd}$ Green's functions for their Bose
symmetric  ones  ${\cal G}_2^{ab}$
and ${\cal Q}_{2\alpha}^{acd}$, holds for the ${\cal S}_{2\alpha}$ term.
It is then easy to show that
\bea
\widetilde\Gamma^{e}_{\!\widetilde A\bar c
c,\alpha}(q)&\equiv&
\Gamma^{e}_{\!A\bar c
c,\alpha}(q)+gf^{eba}\int \!\left\{k_\alpha\left[ 
{\cal G}_1^{ab}(-k+q,k)\right]_{t,{\rm
I}}\right.\nonumber \\
&+&\left.(k-q)_\alpha\left[
{\cal G}_2^{ab}(-k+q,k)\right]_{t,{\rm
I}}\right\}D(-k+q)D(k), \nonumber \\
\widetilde\Gamma^{e}_{\!\widetilde AA\bar c
c,\alpha}(q)&\equiv&
ig^2f^{eba}\int \!\left\{f^{acd}
\left[
{\cal Q}_{2\alpha}^{cdb}(-k+q,k)\right]_{t,{\rm
I}}D(k)\right. \nonumber \\
&- &\left.  f^{bcd}
\left[
{\cal Q}_{1\alpha}^{acd}(-k+q,k)\right]_{t,{\rm
I}}D(-k+q)\right\}.
\eea
This   concludes    the   proof   that   
$\widehat\Gamma^{e}_\alpha\equiv\widetilde\Gamma^{e}_\alpha$.   
We emphasize that
the  sole ingredient in  the above  construction has  been the  STI of
Eq.\r{onshSTI};  in particular,  at  no point  have  we employed  {\it
a priori} the  background 
formalism.  Instead,  its special ghost sector
has  arisen {\it  dynamically},  once the  PT  rearrangement has  taken
place. 

The final step  is to construct the (all  orders) PT gluon self-energy
$\widehat\Pi^{ab}_{\mu\nu}$.   Notice  that at  this  point one  would
expect  that   it  too  coincides  with  the   BFG  gluon  self-energy
$\widetilde\Pi^{ab}_{\mu\nu}$,  since both  the boxes  as well  as the
vertex do coincide  with the corresponding quantities in  BFG, and the
$S$-matrix   is   unique   (renormalization   may   be   carried   out
order-by-order without  any complications, see second paper 
in \cite{Papavassiliou:2000az}). We will carry  out a proof
based on  the strong  induction principle, which  states that  a given
predicate $P(n)$ on $\mathbb N$ is true $\forall\ n\in{\mathbb N}$, if
$P(k)$ is true whenever $P(j)$ is true $\forall\ j\in{\mathbb N}$ with
$j<k$.  We will use  a schematic notation, suppressing Lorentz, color,
and momentum indices.   At one- and two-loop, we  know that the result
is  true  \cite{Cornwall:1982zr,Papavassiliou:2000az}.  Assuming  then
that the PT  construction has been successfully carried  out up to the
order $n-1$,  we will show that  the PT gluon self-energy  is equal to
the  BFG  gluon self-energy  at  order~$n$,  hence  proving that  this
equality holds true at any  given $n$.  From the inductive hypothesis,
we   know   that   $\widehat\Pi^{[\ell]}\equiv\widetilde\Pi^{[\ell]}$,
$\widehat\Gamma^{[\ell]}\equiv\widetilde\Gamma^{[\ell]}$,           and
$\widehat   B^{[\ell]}\equiv\widetilde  B^{[\ell]}\equiv  B^{[\ell]}$,
with $\ell=1,\dots,n-1$. Now, the  $S$-matrix element of order $n$, to
be      denoted      as      $S^{[n]}$,     assumes      the      form
$S^{[n]}=\left\{\Gamma\Delta\Gamma\right\}^{[n]}+B^{[n]}$.    Moreover,
since it  is unique, regardless  if it is  written in the RFG,  in the
BFG, as  well as before and  after the PT rearrangement,  we have that
\mbox{$S^{[n]}\equiv\widehat  S^{[n]}\equiv\widetilde S^{[n]}$}. Using
then  the fact  that $\widehat  B^{[\ell]}\equiv\widetilde B^{[\ell]}$
holds     true     even     when     $\ell=n$,    we     find     that
$\left\{\Gamma\Delta\Gamma\right\}^{[n]}\equiv
\{\widehat\Gamma\widehat\Delta\widehat\Gamma\}^{[n]}\equiv
\{\widetilde\Gamma\widetilde\Delta\widetilde\Gamma\}^{[n]}$.      These
amplitudes can  then be split into  1PR and 1PI  parts; in particular,
the 1PR  part after the PT  rearrangement coincides with  the 1PR part
written            in             the            BFG,            since
$\left\{\Gamma\Delta\Gamma\right\}^{[n]}_{\scriptscriptstyle{\rm   R}}=
\Gamma^{[n_1]}\Delta^{[n_2]}\Gamma^{[n_3]}$  with $n_1,n_2,n_3<n$, and
$n_1+n_2+n_3=n$.   This implies  in turn  the equivalence  of  the 1PI
parts,                 {\it                 i.e.},                 \be
\left(\widehat\Gamma^{[n]}-\widetilde\Gamma^{[n]}\right)
\Delta^{[0]}\Gamma^{[0]}+\Gamma^{[0]}\Delta^{[0]}\left(\widehat\Gamma^{[n]}
-\widetilde\Gamma^{[n]}\right)+                \Gamma^{[0]}\Delta^{[0]}
\left(\widehat\Pi^{[n]}-\widetilde\Pi^{[n]}\right)
\Delta^{[0]}\Gamma^{[0]}\equiv0.  \ee

At this point,
by means of the {\it explicit} construction presented for the
vertex,                  we                  have                 that
$\widehat\Gamma^{[n]}\equiv\widetilde\Gamma^{[n]}$,    so   that   one
immediately gets $\widehat\Pi^{[n]}\equiv\widetilde\Pi^{[n]}$.  Hence,
by  strong  induction,  the  above  relation is  true  for  any  given
perturbative      order     $n$,      {\it     i.e.},      we     have
$\widehat\Pi^{ab}_{\mu\nu}\equiv\widetilde\Pi^{ab}_{\mu\nu}$, {\it q.e.d.}

In conclusion, we have shown that the use of the STI \r{onshSTI}
satisfied by the special Green's function \r{sgf}, allows for the
generalization of the PT procedure to all orders. 
It would be interesting to further explore the physical meaning of 
the $n$-point functions obtained \cite{Bernabeu:2002nw}, 
and establish possible connections with related 
formalisms \cite{Rebhan:1984bg}.

{\it Acknowledgments:} This work has been supported by the CICYT Grants
AEN-99/0692 and BFM2001-0262. D.~B. thanks the Theoretical
Physics Department of the University of Trento, where part of this work
has been carried out.

\end{document}